\documentclass[pra,aps,showpacs,articles,nofootinbib,twocolumn]{revtex4-1}
\usepackage{epsf,epsfig}
\usepackage[psamsfonts]{amssymb}
\usepackage{amsmath}
\usepackage{color}

\def\be{\begin{equation}}
\def\ee{\end{equation}}
\def\ba{\begin{eqnarray}}
\def\ea{\end{eqnarray}}

\begin{document}

\title{Measurement enhances long-distance Entanglement generation in spin chains with dissipative processes}
\author{Morteza Rafiee}

\affiliation{Department of Physics, Shahrood University of Technology, 3619995161 Shahrood, Iran}

\date{\today}

\begin{abstract}
	In this paper, effects of the regular measurements on a noisy channel has been investigated. The strategy introduced by A. Bayat, and Y. Omar [New J. Phys. {\bf 17}, 103041 (2015)] is followed to suppress  dephasing and dissipation effects in a noisy spin channel and generate long distance entanglement by global measurement on the channel.  A regular global measurements performed on spin channel weakly coupled to the sender and receiver qubits via $XX$ interaction. This scheme is applied for the dephasing and dissipation in non-zero temperature processes separately and the results show that amounts of achieved entanglement enhanced rather than the no-measurement approach.
\end{abstract}

\maketitle

\section{Introduction}
It is well understood that entanglement, i.e. non-local quantum correlations \cite{Schroedinger} is the key resource for quantum communication and computation
\cite{QCC} and has been verified in such protocols as cryptography \cite{crypt} and teleportation \cite{telep}.
Mediating interaction between arbitrary pairs in a network of qubits is essential for realizing two-qubit entangling gates. Unlike the neighboring qubits, interacting the distant ones is notoriously difficult. One way to mediate interaction between distant qubits is to use an additional setup, called quantum bus. Among the numerous quantum
systems suitable for quantum bus and quantum networks implementation, spin chains are the most common buses where their tunable interaction has motivated researchers to use this permanent potential in the information processes \cite{bose, giovannetti, osborne, bayat3, bayat2}. However, quantum systems would unavoidably interact with their environments by
dissipative processes which leads to fragile entanglement. Nevertheless, there are a variety of entanglement preserving
mechanisms that have been put forward \cite{preserv,Julsgaard,ss-ent-p,ss-ent,Polzik, BKraus, Memarzadeh, Rafiee_Lupo, Ghanbari_Rafiee, nourmandi_Rafiee}.
Moreover, to overcome the loss of entanglement over the distance in spin chains, one has to either delicately engineer the couplings \cite{Yung-bose,bayat-gate} or switch to super slow perturbative regimes \cite{Lukin-gate,WojcikLKGGB}.
Alternatively, one may use intermediate spins as interaction mediators between a sender and a receiver and use the rotational protocol and single projective measurement to create long distance entanglement \cite{Rafiee_proj}. It was shown that, projective measurements are essential elements in some quantum technologies, such as entanglement generation between superconducting qubits within a meter of distance \cite{Roch}, entangling macroscopic atomic ensembles \cite{pederson} and many-body state engineering \cite{sherson}. Moreover, measurements protocols has been used as a mean for transport as well as small couplings in different protocols \cite{shizume, Gualdi, pouyandeh, bayat_d} and also using the global measurements, performed regularly on the spin channel, can suppress the effect of dephasing of the state transfer which has been investigated in \cite{bayat_omer}  \\
The most common approach to realizing an effective long-distance coupling is to use a
quantum mediator, as has been demonstrated in quantum dot arrays \cite{Hanson, Veldhorst}, superconducting qubits \cite{Majer, Sillanpaa} and in trapped ions \cite{Schmidt}. Many approaches to implementing coherent spin coupling between
distant quantum dots have been proposed using a variety of coupling mechanisms. These include coherent spin-exchange \cite{Baart}, superconductors \cite{Leijnse, Hassler} and superexchange via intermediate quantum dots \cite{Mehl}.
long distanse entanglement via spin chain has been investigated both theoretically \cite{Venuti_1, Venuti_2, Rafiee_proj} and exprimentally, \cite{Trifunovic, Sahling}.  
However, recent advances in fabrication of quantum dot arrays \cite{puddy} and also the experimental realization of the Mott insulator phase for both bosons \cite{boson} and fermions \cite{fermion}, with exactly one atom per site, in optical lattices has lead to possible effective spin Hamiltonians \cite{Lukin}. 

In this paper, we put forward the approach in Ref.\cite{bayat_omer} to investigate entanglement generation between the ends of a spin chain which its dynamics is governed by $XX$ Hamiltonian. The intermediate qubits between the ends of the chain, spin channel, interact with their baths and their dynamics is effected by dephasing or dissipation. Our protocol, based on performing regular global measurements on this channel leads to compensation of dissipative process effects.

The structure of this paper is as follows: in section (\ref{sec2}) we introduce our setup. In section (\ref{sec3}) our setup has been analyzed without any dissipative processes
and effects of dissipation and our measurement protocol on entanglement generation is investigated for a noisy channel in both measurement and no-measurement approach in section (\ref{sec4}). We finally summarize our results in section (\ref{sec5}).

\section{set-up}\label{sec2}

We consider a chain of $N$ spin $1/2$ particles, where $N$ is even, interacting through a $XX$ Hamiltonian. The spins at the ends of the chain are weakly interacting with the intermediate spins which spins labeled from 2 to $N-1$. These intermediate spins play the role of our spin channel and its Hamiltonian is

\begin{equation}\label{hamiltonian}
H_{ch}=J\sum_{k=2}^{N-1}[\sigma_k^+ \sigma_{k+1}^- + \sigma_k^- \sigma_{k+1}^+]
\end{equation}
where J is the exchange coupling and $\sigma_k^\pm$ are the Pauli spin ladder operators acting on site k.
The spins at sites $1$ and $N$ which are sender and receiver qubits, interact weakly with the channel as
\begin{equation}\label{hamiltonian}
H_{I}=J'(\sigma_1^+ \sigma_2^- + \sigma_1^- \sigma_2^+ + \sigma_{N-1}^+ \sigma_N^- + \sigma_{N-1}^- \sigma_N^+)
\end{equation}
Where $J'$ is the coupling of the encoding qubits with channel. Hamiltonian of the total system is described by $H=H_I+H_{ch}$. 
The state $|+ \rangle=\frac{1}{\sqrt{2}}(|0 \rangle + |1 \rangle )$ encode on both the qubits $1$ and $N$ while the channel is initialized in the ferromagnetic state $|0_{ch}\rangle=|0,0,...,0\rangle$. So, the initial state of the whole system can be written as
\begin{equation}\label{hamiltonian}
\rho(0)=|+\rangle \langle +| \otimes |0_{ch} \rangle \langle 0_{ch}| \otimes |+\rangle \langle +|.
\end{equation}
Due to the interaction of the channel with its environment, dynamics of the system is governed by the master equation \cite{Breuer}
\begin{equation}\label{master_eq}
\dot{\rho}=-\iota\left[H,\rho\right]+(\bar{n}+1){\cal D}\left[L\right]\rho + (\bar{n}) {\cal D'}\left[L\right]\rho
\end{equation}
\begin{eqnarray}
\mathcal{D}(\rho) = \frac{\gamma}{2}
\left( 2 L_k \rho L_k^\dag - L_k^\dag L_k \rho - \rho L_k^\dag L_k \right), \nonumber \\ 
\mathcal{D'}(\rho) = \frac{\gamma}{2}
\left( 2 L_k^\dag \rho L_k - L_k L_k^\dag \rho - \rho L_k L_k^\dag \right), \nonumber \\ 
\end{eqnarray}
with $\gamma>0$ and
\begin{equation}
L_k = \sigma_k^-, \hspace{5mm} k=2,3,...,N-1 
\end{equation}
The non-Hamiltonian term $\mathcal{D}(\rho)$ describes a Markovian damping process
in which qubits coherently decay into the nonzero temperature bosonic bath with decay rates $\gamma$ and $\bar{n}$ is the mean number of bath quanta.
Another non-desire effect can be considered as dephasing and the evolution of the system under the dephasing, can be described by the following master equation \cite{Breuer}
\begin{eqnarray}\label{me}
\dot{\rho}=-\iota\left[H,\rho\right]+{\cal D}\left[L\right]\rho,
\end{eqnarray}
where
\begin{eqnarray}
\mathcal{D}(\rho) = \frac{\gamma}{2}
\left( 2 L_k \rho L_k^\dag - L_k^\dag L_k \rho - \rho L_k^\dag L_k \right), \nonumber \\  
\end{eqnarray}
with $\gamma>0$ and
\begin{equation}
L_k = \sigma_k^z, \hspace{5mm} k=2,3,...,N-2 
\end{equation}
 Random level fluctuations in the system of trapped atoms in the optical lattice due to the fluctuation of surrounding magnetic or electric fields can be considered as dephasing. 

This $XX$ Hamiltonian can be realized in experiment and a physical realization of it is the setting of ultracold atoms trapped in an optical lattice. An optical lattice made of a standing wave formed by two different sets of laser beams. The resulting potential is
 
 \begin{equation}\label{V_lattice}
 V(x)=V_l \cos^2(2\pi x/ \lambda_l)+V_s \cos^2(2\pi x/\lambda_s)
 \end{equation}
 
 where, $\lambda_l=2\lambda_s$ are the wave lengths, $V_l$ and $V_s$ are the amplitudes. The low energy Hamiltonian of atoms trapped by $V(x)$ is \cite{Lukin}
 
 \begin{align}\label{Hubbard}
 	H=&-\sum_{<i,j>,\sigma} (J_{i\sigma} a^\dagger_{i,\sigma} a_{j,\sigma}+H.C.) + U_{\uparrow \downarrow} \sum_i n_{i,\uparrow}n_{i,\downarrow} \nonumber \\
 	&+ \frac{1}{2}\sum_{i,\sigma}U_{\sigma}n_{i,\sigma}(n_{i,\sigma}-1) ,
 \end{align}

 where, $<i,j>$ denotes the nearest neighbor sites, $a_{i,\sigma}$ annihilates one atom with spin $\sigma=\uparrow,\downarrow$ at site $i$, and $n_{i,\sigma}=a^\dagger_{i,\sigma} a_{i,\sigma}$. We are interested in the regime where $J_i\ll U_{\sigma},U_{\uparrow \downarrow}$. This choice of hopping terms energetically prohibit the multiple occupancy of any site which corresponds to an
 insulating phase. The effective Hamiltonian is found to be\cite{Lukin, clark}
 
 \begin{eqnarray}\label{XXZ Hamiltonian}
 H&=&\sum_{<i,j>} J_i^{z}\sigma_i^z\sigma_j^z - \sum_{<i,j>} J_i^{\perp}(\sigma_i^x\sigma_j^x+\sigma_i^y\sigma_j^y),
 \end{eqnarray}
 
 where$\sigma_i^x=a^\dagger_{i\uparrow}a_{i\downarrow}+a^\dagger_{i\downarrow}a_{i\uparrow}$ and $\sigma_i^y=-i(a^\dagger_{i\uparrow}a_{i\downarrow}-a^\dagger_{i\downarrow}a_{i\uparrow})$
 are the pauli's spin operators. The effective couplings $J_i^z$ and $J_i^{ex}$ are given by
 
 \begin{equation}
 J_i^{z}=\frac{J_{i\uparrow}^2+J_{i\downarrow}^2}{2U_{\uparrow \downarrow}}-\frac{J_{i\uparrow}^2}{U_{\uparrow}}-\frac{J_{i\downarrow}^2}{U_{\downarrow}}, \hspace{10mm} J_i^{\perp}=\frac{J_{i\uparrow}+J_{i\downarrow}}{U_{\uparrow \downarrow}},
 \end{equation}
 
 The optical lattice parameters could be engineered such that $U_{\uparrow}=U_{\downarrow}=2U_{\uparrow\downarrow}=U$ and $J_{i\uparrow}=J_{i\downarrow}=J_i$. So the effective Hamiltonian is reduced to the $XX$ spin Hamiltonian \cite{Lukin, clark}.
 
 \begin{eqnarray}\label{XX Hamiltonian}
 H&=&-\sum_{<i,j>} J_i^{\perp}(\sigma_i^x\sigma_j^x+\sigma_i^y\sigma_j^y),
 \end{eqnarray}
 
 The tunneling $J_i$'s are controlled by the amplitudes $V_l$ and $V_s$\cite{Lukin}. In this system we also consider $J_1=J_N=j'$ and $J_k=J$ for $k=2,3,...,N-1$.
  
  \section{no dissipative processes}\label{sec3}
  In the case of no dissipative processes, where $\gamma=0$, above Hamiltonian can be reduced to an effective Hamiltonian \cite{bayat_omer} as follows
  
  \begin{equation}\label{hamiltonian_effective}
  H_e=J_e(\sigma_1^+ \sigma_N^- + \sigma_1^- \sigma_N^+)
  \end{equation}
  where
  \begin{equation}\label{hamiltonian_effective}
  J_e=(-1)^{\frac{N}{2}}\frac{J'^2}{J}.
  \end{equation}
  
  this result is obtained by the condition $J'<<J$. We will consider this coupling strength regime in this paper.
  The encoding information at the ends of the chain will be entangled via $XX$ Hamiltonian.
  Entanglement between the ends of the chain at each time can be obtained by calculating the concurrence\cite{wooters}
  
\begin{equation}
C(t)=max\{0,2\lambda_{max}(t)-\sum_{i=1}^4\lambda_i(t)\},
\end{equation}
where $\lambda_i$ are the eigenvalues of the matrix $\sqrt{\rho_{1N} \sigma^y \otimes \sigma^y \rho_{1N}^* \sigma^y \otimes \sigma^y}$ while $\rho_{1N}$ is the reduced density matrix of the qubits $1$ and $N$.
  \begin{figure} \centering
	\includegraphics[width=9cm,height=5.5cm,angle=0]{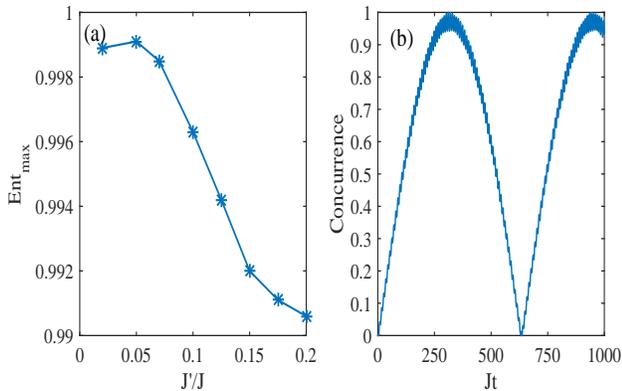}
	\caption{(Color online) (a) Maximum values of the entanglement between the ends of the spin chain with $N=10$ spins versus the $J'/J$  while $\gamma=0$. (b) Dynamics of the entanglement between the ends of the same spin chain versus time while $J'=0.05J$.}\label{gamma0}
\end{figure}
 In order to determine the optimum value of $J'$, maximum values of the entanglement between ends of the spin chain with $N=10$ spins for different values of the ratio $J'/J$ has been plotted in Fig(\ref{gamma0})-(a) while $\gamma=0$. As can be seen from these results, the suitable values for the ratio $J'/J=0.05$ and we will use $J=1$ and $J'=0.05$ for the following calculations. Moreover, dynamics of this entanglement versus the time for the same spin chain has been shown in Fig(\ref{gamma0})-(b). 


\section{Effect of dissipation and regular measurement}\label{sec4}
\begin{figure} \centering
	\includegraphics[width=9cm,height=5.5cm,angle=0]{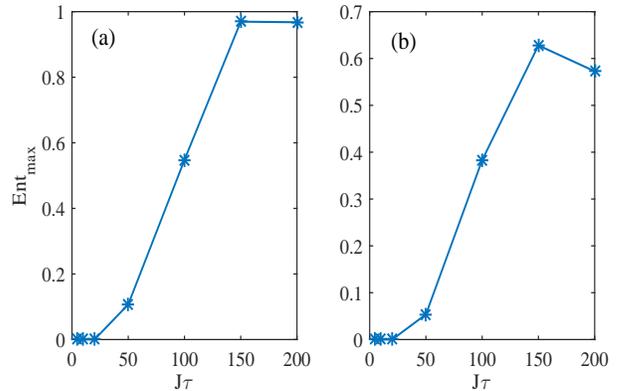}
	\caption{(Color online) Maximum values of the entanglement between the ends of the spin chain with $N=10$ spins and $J'=0.05J$ versus the $J\tau$  while (a)$\gamma=0$, (b)$\gamma=0.05$ and $\bar{n}=0.05$}\label{E_max_tau}
\end{figure}

\begin{figure} \centering
	\includegraphics[width=9cm,height=6cm,angle=0]{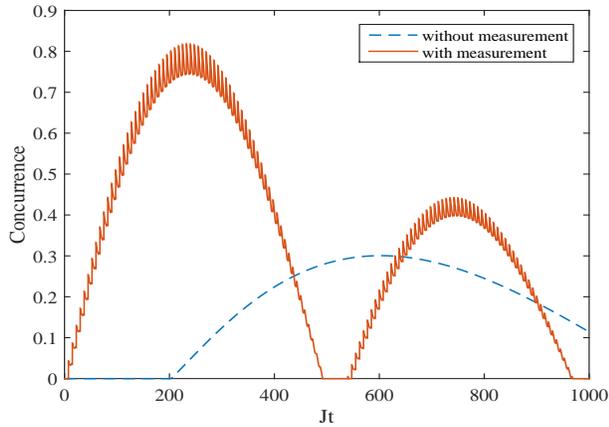}
	\caption{(Color online) Dynamics of the entanglement between the ends of the spin chain with $N=10$ spins versus time while $J'=0.05J$, $\gamma=0.02J$, $\bar{n}=0.05$ and $\tau=150/J$ for performing measurement}\label{N10_gamma02n05}
\end{figure}

\begin{figure} \centering
	\includegraphics[width=9cm,height=5.5cm,angle=0]{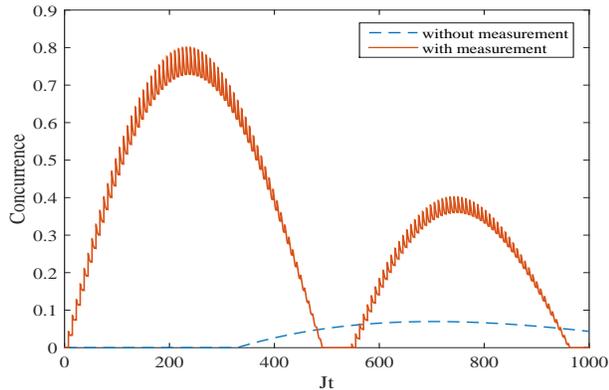}
	\caption{(Color online) Dynamics of the entanglement between the ends of the spin chain with $N=10$ spins versus time while $J'=0.05J$, $\gamma=0.02J$, $\bar{n}=0.1$ and $\tau=150/J$ for performing measurement}\label{N10_gamma02n1}
\end{figure}

\begin{figure} \centering
	\includegraphics[width=9cm,height=6cm,angle=0]{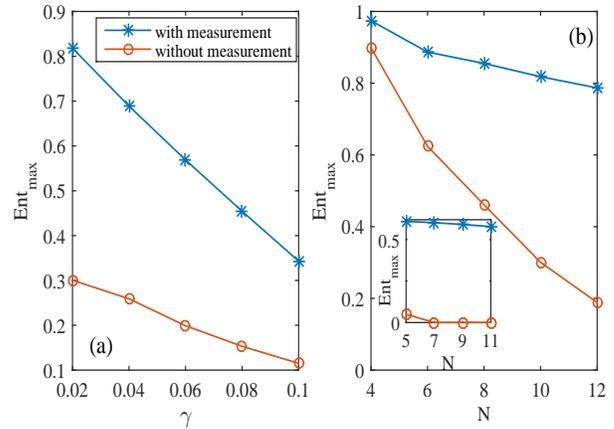}
	\caption{(Color online) (a) Maximum values of the entanglement between the ends of the spin chain with $N=10$ spins versus decay rate $\gamma$ (b) Maximum values of the entanglement for a chain with $\gamma=0.02$ versus even $N$ (inset) versus odd $N$ while $J'=0.05J$, $\bar{n}=0.05$ and $\tau=150/J$ for performing measurement.}\label{Ent_gamma}
\end{figure}

 In the presence of dissipation, dynamics of the system is governed by the Eq(\ref{master_eq}). Here, excitations of the encoding qubits leak to the dissipative channel and so lead to decrease the amounts of entanglement. Following the similar protocol introduced in \cite{bayat_omer} for state transfer, we introduce the regular measurements on the channel to preserve the leakage of excitations.
  The corresponding projection operators are 
 \begin{equation}
 M_0=|0_{ch}\rangle \langle 0{ch}|, \hspace{5mm} M_1=I-M_0,
 \end{equation}
 where $I$ is identity operator. The projective measurements are performed regularly on the channel, at time intervals $\tau$. For the very small amounts of $\tau$ dynamics is frozen as predicted by the Zeno effect \cite{zeno}. The border of the Zeno and non-Zeno regime happens when $\tau \sim 1/J'$ \cite{bayat_omer}. So, to avoid the Zeno effect we should choose the larger time intervals. For the sake of clarify, maximum amount of entanglement for a chain with $N=10$ and in the case of with $\gamma=0$ (resp. $\gamma=0.05$ and $\bar{n}=0.05$), has been plotted versus the time interval $J \tau$ in Fig. (\ref{E_max_tau})-(a),(b). As can be seen from the results, $J\tau=150$ is the optimum value which will be used in the following calculations and also is comparable with the values used in \cite{bayat_omer}.
Furthermore, in order to show the supremacy of our strategy, in Figure (\ref{N10_gamma02n05}), we have plotted the entanglement dynamics for a chain with $\gamma=0.02J$, $J'=0.05J$, $N=10$ and $\bar{n}=0.05$ while projection measurement on the channel was performed with $\tau=150/J$ (Solid line) and also the same dynamics without any measurement (dashed line). Amount of $\bar{n}=0.05$ is according to $T \approx 33 mK $ \cite{Tsomokos}. Moreover, the results of the same calculations have been done for another temperature $\bar{n}=0.1$ which is according to $T \approx 41 mK $ \cite{Tsomokos} and have been shown in Fig. (\ref{N10_gamma02n1}). These results demonstrate the maximum amount of entanglement can be increased via the projection measurements. However, superiority of this measurement protocol is obvious for larger temperature by comparison of the differences of the entanglement with and without measurement in both Figs. (\ref{N10_gamma02n05}) and (\ref{N10_gamma02n1}).
In addition, maximum amount of entanglement has been plotted as a function of $\gamma$ in Fig. (\ref{Ent_gamma}) (a) for a chain with $N10$ and also maximum values of the entanglement with damping rate $\gamma=0.02$ has been plotted versus the number of spins even $N$ (resp. odd $N$) in Fig. (\ref{Ent_gamma}) (b) (resp. inset) while $J'=0.05J$, $\bar{n}=0.05$ and $\tau=150/J$ for both subplots. These figures show that the maximum amounts of entanglement decay by increasing $\gamma$ and $N$ while measurement approach enhances entanglement generation for all ranges of them. Moreover, although the channel with odd number of spin does not work for entanglement generation under the dissipation processes (inset of Fig. (\ref{Ent_gamma}) (b)), these regular measurements bring them to be useful for this task.    

\begin{figure} \centering
	\includegraphics[width=9cm,height=6cm,angle=0]{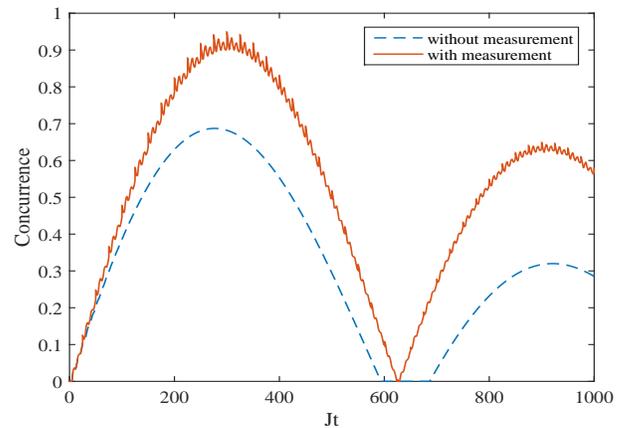}
	\caption{(Color online) Dynamics of the entanglement between the ends of the chain with $N=10$ versus time while $J'=0.05J$, $\gamma=0.02J$ and $\tau=500/J$ for performing measurement}\label{SzN10_gamma02}
\end{figure}

To finalize our analysis we also study the performance of
our protocol on a dephasing channel has been considered and entanglement of the first and last qubits has been plotted versus time $t$ in Fig. (\ref{SzN10_gamma02}) for the case of $N=10$, $\gamma=0.02J$, $J'=0.05J$ and $\tau=500/J$ while here $\bar{n}=0$ and $L_k=\sigma_k^z$. Moreover, these results with large $\tau$ show that we do not need to perform several measurement protocols during the evolution for this system.    


\section{Conclusion}\label{sec5}
 In this paper we exploit the effects of regular global measurement on a noisy channel to suppress the effects of dissipative processes. In fact, these measurements enhance the entanglement generation between sender and receiver qubits which are weakly coupled to the noisy channel and to avoid the Zeno effect, we do not need to perform many measurement during the evolution of the system. In fact, our measurement protocol offers much higher entanglement generation than the traditional no-measurement approach in both non-zero temperature dissipation and dephasing effects. Moreover, our scheme is superior in systems which suffer the dissipation with larger temperature.  
 
\section{Acknowledgments}
The author thanks A. Bayat for useful discussion and comments at the University College London.

\appendix

\end{document}